\documentclass[aps,pre,reprint,superscriptaddress,floatfix,a4paper]{revtex4-1}
\usepackage{graphicx}
\usepackage{epsf}
\usepackage{caption}
\usepackage{subfig}
\usepackage{hyperref}
\usepackage{epstopdf}

\begin{document}
\newcommand{\newc}{\newcommand}
\newc{\mbf}{\mathbf}
\newc{\boma}{\boldmath}
\newc{\beq}{\begin{equation}}
\newc{\eeq}{\end{equation}}
\newc{\beqar}{\begin{eqnarray}}
\newc{\eeqar}{\end{eqnarray}}
\newc{\beqa}{\begin{eqnarray*}}
\newc{\eeqa}{\end{eqnarray*}}
\newc{\bd}{\begin{displaymath}}
\newc{\ed}{\end{displaymath}}


\title{Surface properties and scaling behavior of a generalized 
ballistic \\ deposition model in $(1+1)$-dimension}
\author{Baisakhi Mal} 
\email{baisakhi.mal@gmail.com}
\affiliation{Department of Physics, Jadavpur University, Calcutta
700 032, India.}
\affiliation{Department of Physics, Budge Budge Institute of Technology, Calcutta
700 137, India.}
\author{Subhankar Ray}
\email{sray.ju@gmail.com, sray@phys.jdvu.ac.in}
\affiliation{Department of Physics, Jadavpur University, Calcutta
700 032, India.}
\author{J. Shamanna}
\email{jlsphy@caluniv.ac.in}
\affiliation{Physics Department, University of Calcutta, Calcutta 700 009, India.}
\begin{abstract}
The surface exponents, the scaling behavior and the bulk porosity 
of a generalized ballistic deposition (GBD) model are studied. 
In nature, there exist particles with varying degrees of 
stickiness ranging from completely non-sticky to fully sticky. 
Such particles may adhere to any one of the successively encountered
surfaces, depending on a sticking probability
that is governed by the underlying stochastic mechanism.
The microscopic configurations possible in this model are much
larger than those allowed in existing models of ballistic deposition
and competitive growth models that seek to mix ballistic and random
deposition processes.
In this article, we find the scaling
exponents for surface width and porosity for the proposed GBD model.
In terms of scaled width $\widetilde{W}$ and scaled time $\tilde{t}$, 
the numerical data collapse on to a single curve, demonstrating 
successful scaling with sticking probability $p$ and system
size $L$. Similar scaling behavior is also found for the porosity.
\end{abstract}
\date{\today}
\maketitle

\section{Introduction}
The formation and growth of rough surfaces have several applications in 
physical and chemical processes, such as
crystal growth, growth of thin films, vapor deposition, formations of colloids
and electroplating to name a few
\cite{bara95, mea93}.
Many of the unique mechanical, optical and electromagnetic properties 
of surfaces originate from their surface morphology.
The bulk properties of depository rocks, e.g., 
porosity, saline saturation, texture, stability and strength,
find important applications in geology of sedimentary rocks.
The underlying formation mechanism influences the
geometry of deposition structures and 
is relevant in the manufacture of optical and electronic nanostructures and
nanodevices, sophisticated drug delivery systems using magnetic carbon
nanostructures \cite{med1} and smart nanostructures for monitoring, diagnoses and 
treatment in physiology \cite{med2}.
Understanding the dynamics and growth of surfaces is, therefore, a  
challenging problem in surface science. 

There are two fundamental approaches to the study of surface growth: (i) 
extensive numerical simulation of discrete models and computation of 
surface and bulk properties \cite{mea93}; and (ii) solving the stochastic 
differential equation derived from phenomenology corresponding to 
the growth model \cite{ew82,kpz86}.
Another recent approach of study involves transformation of discrete 
deposition rules into stochastic differential equations using a limiting
procedure and regularization, and hence, finding the scaling exponents \cite{vve03a}.

\begin{figure}
\subfloat[Random deposition]{\includegraphics[width=2.4cm]{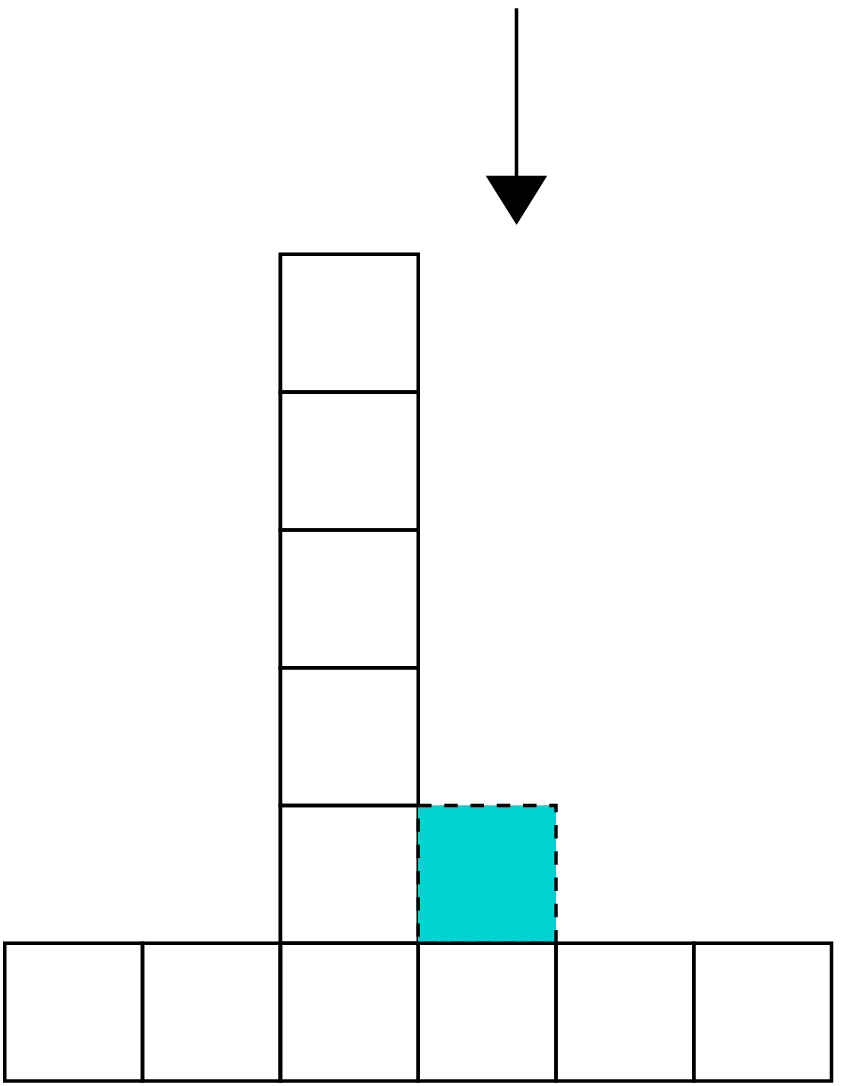}
  \label{RD}}%
  \qquad
  \subfloat[Ballistic deposition]{\includegraphics[width=2.4cm]{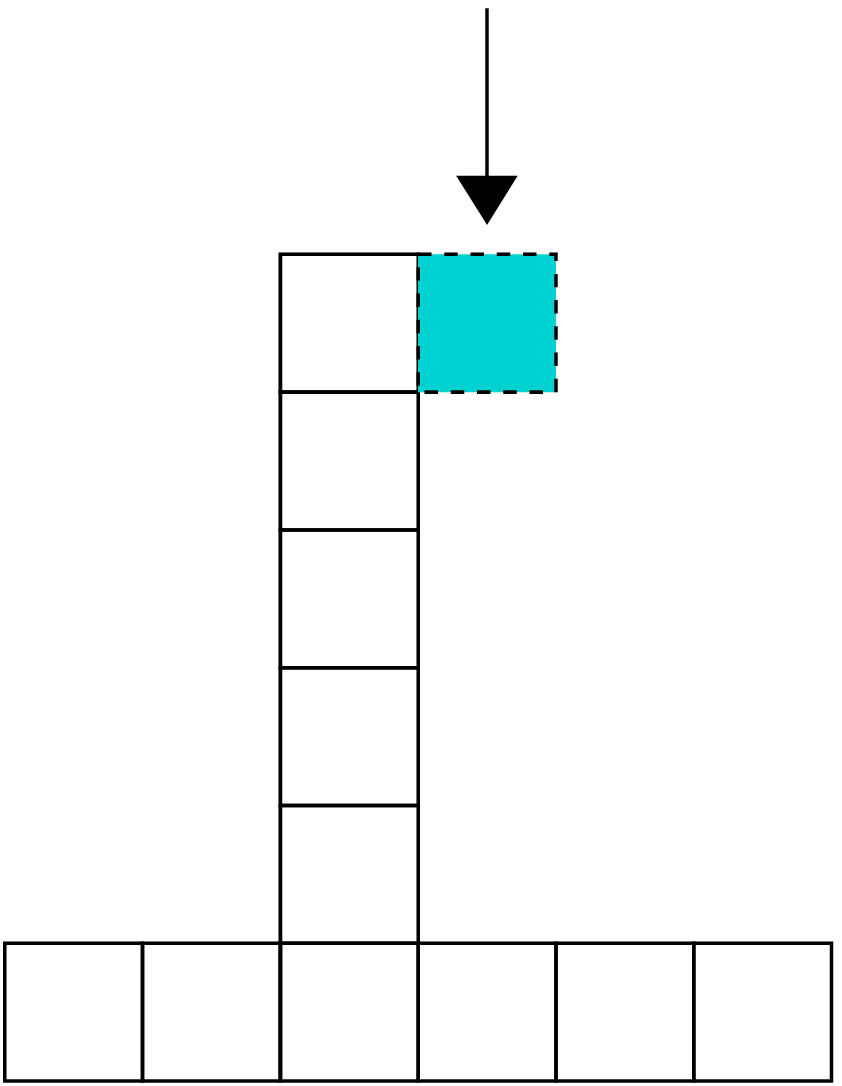}
  \label{BD}} \\%
  \subfloat[][HABD]{\includegraphics[width=2.3cm]{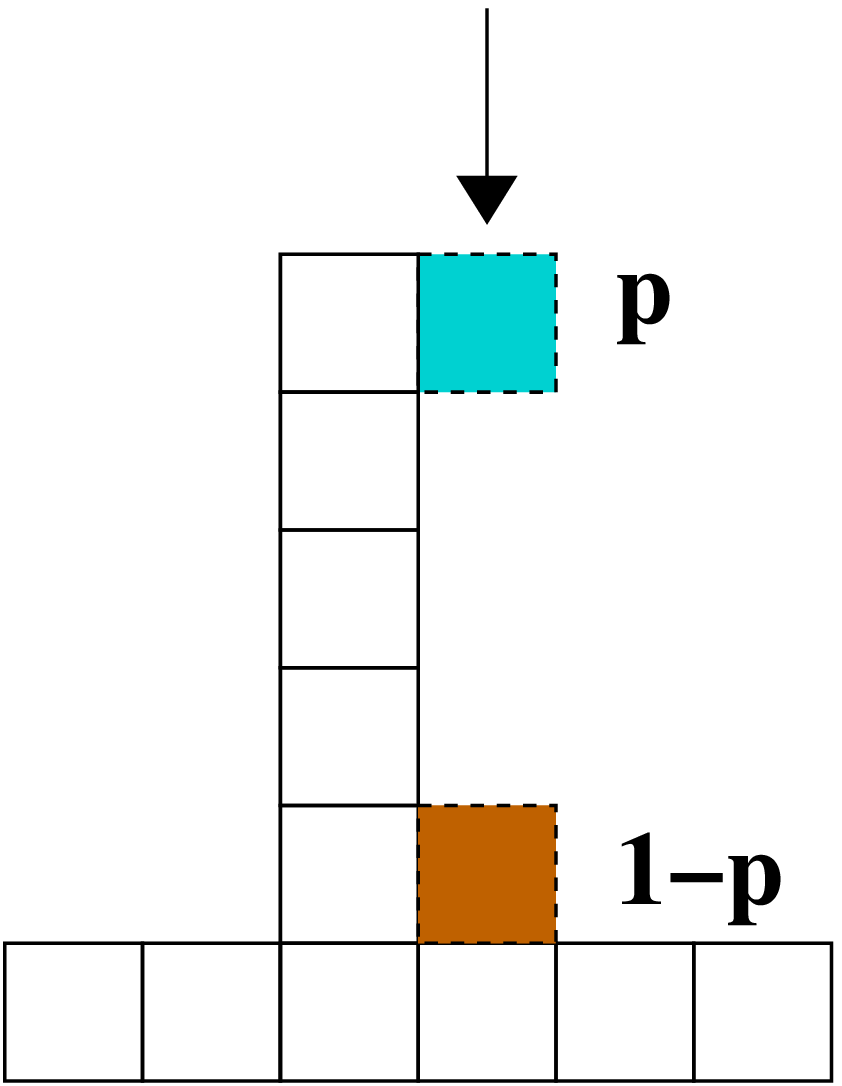}%
  \label{horoBD}}%
  \qquad
  \subfloat[][HABD1]{\includegraphics[width=2.3cm]{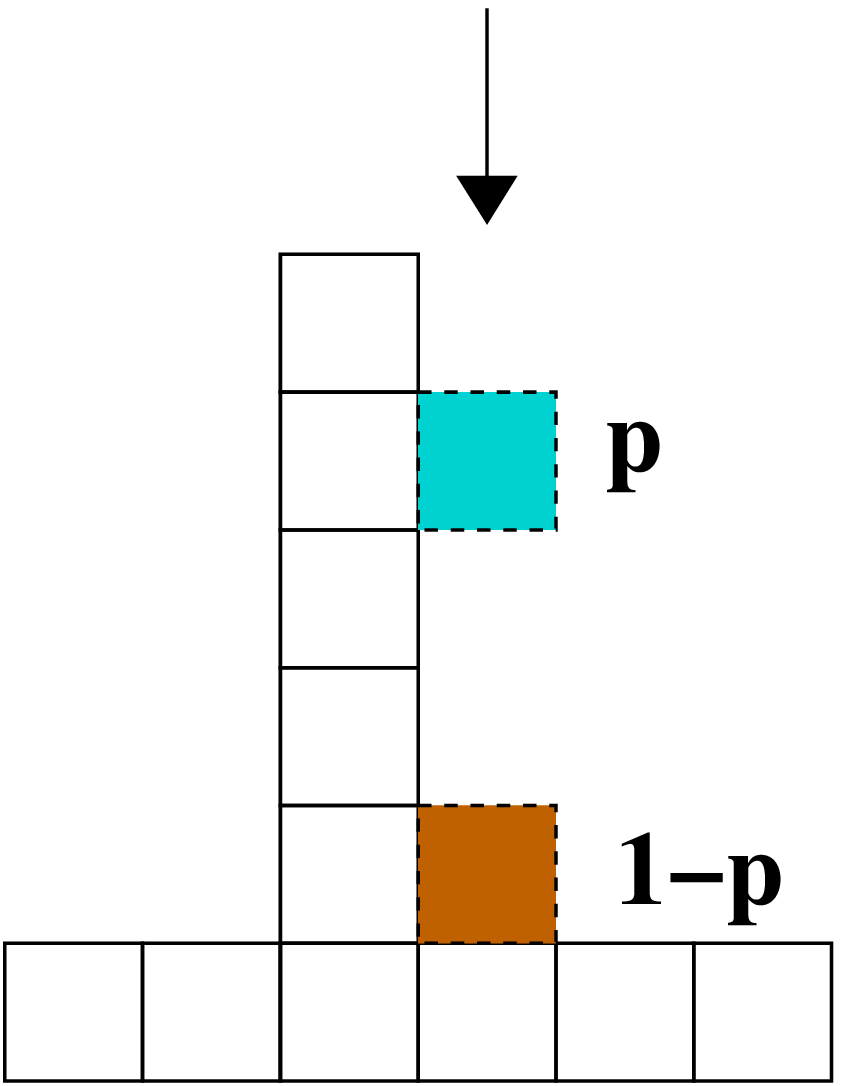} \label{horoBD1}} 
  \qquad
  \subfloat[][HABD2]{\includegraphics[width=2.3cm]{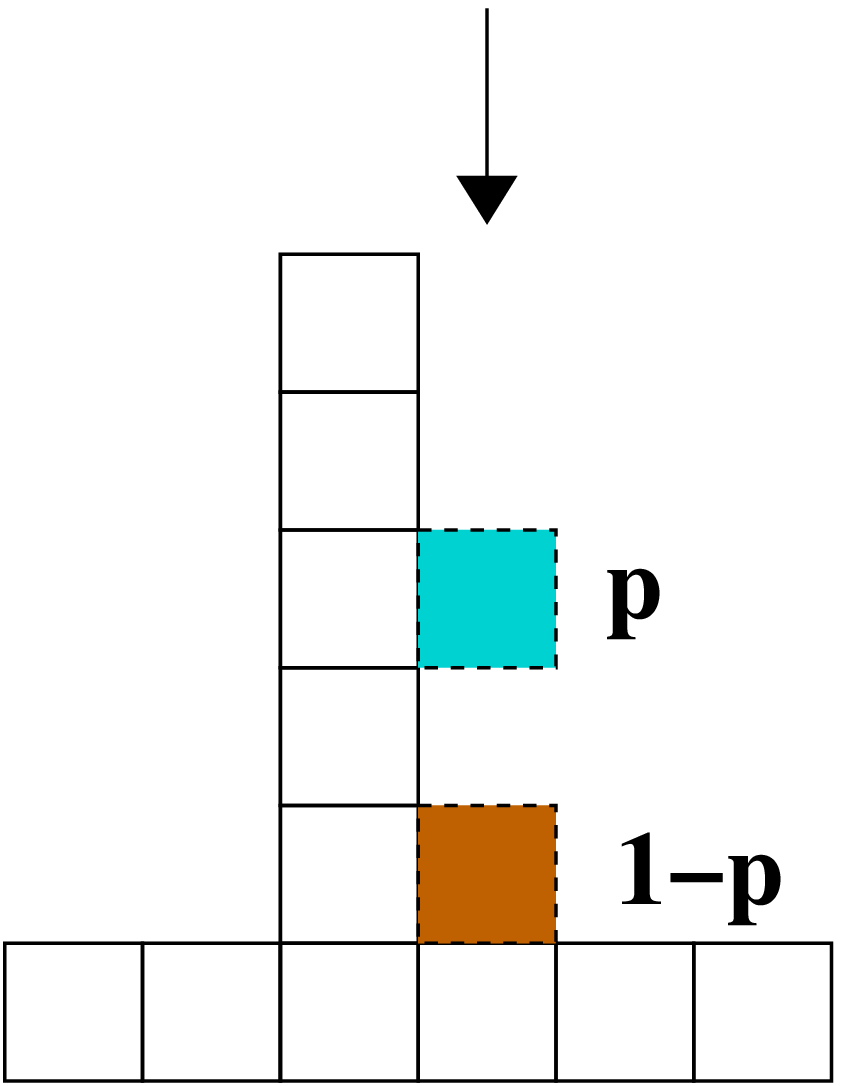} \label{horoBD2}}\\%
  \subfloat[present model (GBD)]{\includegraphics[width=4.5cm]{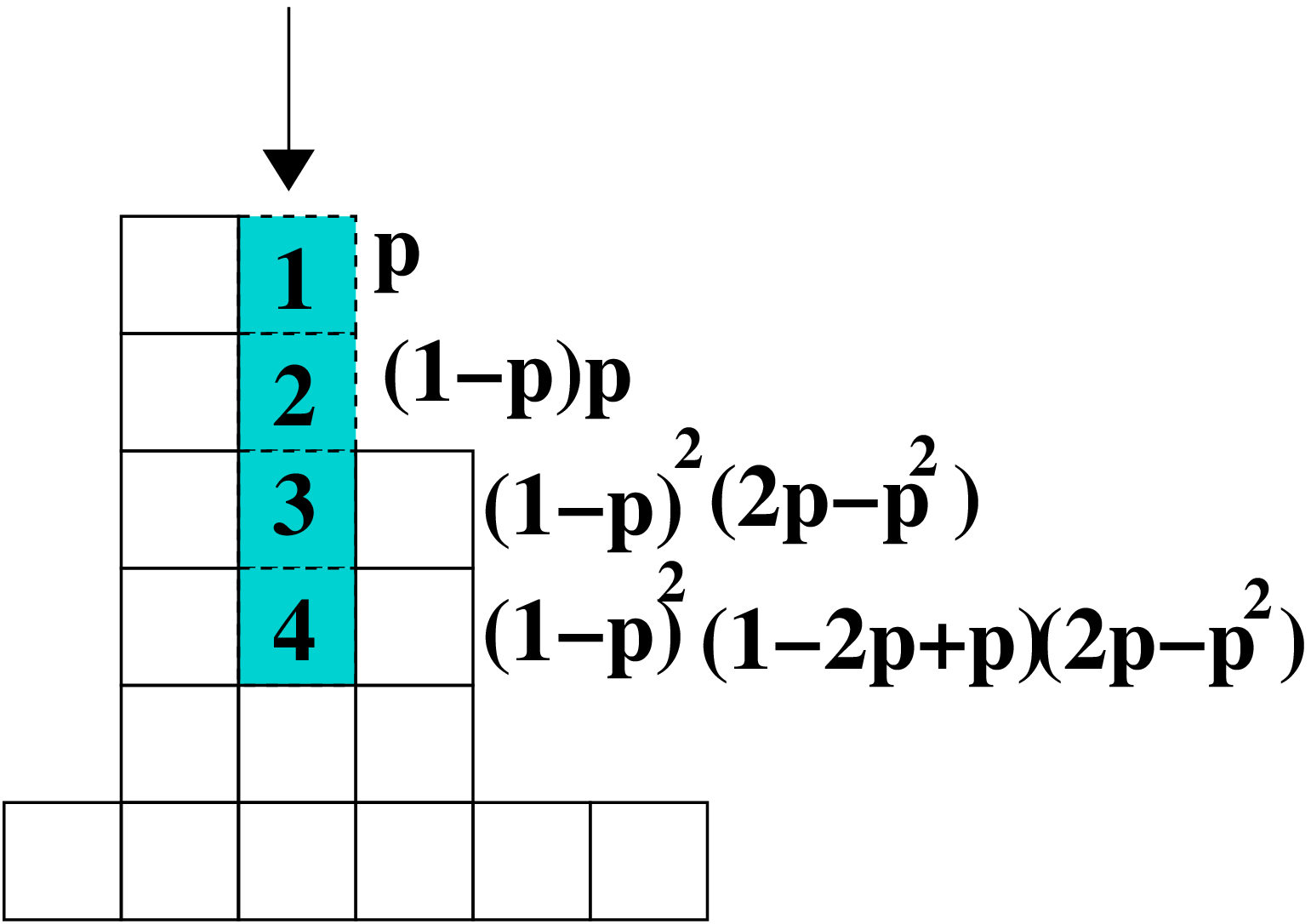}
    \label{gRDBD}}%
    \caption{(Color Online) Allowed positions of the deposited particle for 
    (a) RD, (b) BD, \\ (c, d, e) Horowitz and Albano's BD appearing in \cite{horo01b,horo06}, HABD, HABD1, HABD2, \\ (f) present model (GBD). 
      \label{genRDBD}}
\end{figure}

Random deposition (RD) is a simple deposition process where non-sticky, 
solid particles deposit on randomly selected sites of a substrate
 (see Fig \ref{RD}).
A quantitative measure of the roughness of the surface, called
the surface width $W(L,t)$, is defined in terms of the surface height 
$h(i,t)$, at a site $i$ and at a time $t$, as,
\beq\label{surfwid}
W(L,t) = \sqrt{ \frac{1}{L} \sum_{i=1}^{L} \left[ h(i,t) -
\langle h(t) \rangle \right]^2 } ,
\eeq
where $L$ is the system size, and $\langle \;\; \rangle$ is the average.
In RD, each site grows independently of
the other sites and the surface roughness grows without bound. 
There are no voids in random deposition, so the deposition 
structure is compact \cite{bara95}.

Ballistic deposition (BD) on the other hand, gives rise to porous 
structures as the depositing particles stick to the first surface they
encounter in their vertical downward journey towards randomly 
selected sites (see Fig \ref{BD}). In BD, the particles behave
as strongly sticky, whereas in RD they are 
not sticky at all \cite{bara95, fam85,fam90}.
In nature however, particles may have intermediate stickiness
which varies between the two extremes of strongly sticky and
completely non-sticky behavior. 

In the present study of a generalized ballistic deposition (GBD)
model, we investigate the deposition of physically realistic 
particles with intermediate stickiness. The level of stickiness is 
parametrized in terms of a sticking probability $p$ of the 
incident particle, at each contact point with the surface. 
The parameter $p$ ranges between $0$ and $1$, $0$ representing 
non-sticky particles as in the case of random deposition, 
and $1$ representing extremely sticky particles, as in ballistic
deposition.
 A logarithmic plot of the surface width $W(L,t)$ versus $t$, for GBD
shows three distinct growth regions, similar to those observed by 
Banerjee et. al.\cite{kas14}, followed by saturation.
The present model, for any $p>0$, however small, 
leads to porous structures \cite{katzav2006}. 
Interesting scaling relations of surface roughness and porosity, 
with both the system size $L$, and the sticking probability $p$,
are observed in the growth \cite{kpz86} and saturation regions.
The relevant scaling exponents are determined from the simulation data.

This GBD is distinct from the earlier competitive growth 
models \cite{horo01a,horo01b,horo06, wang93, wang95, nashar00, 
reis02, reis03,reis06,horo06,pell90,pell91} in several ways. 
The deposition in GBD mimics a realistic sticking process such as gel, or 
mud thrown on a wall. 
Some variants of BD models, e.g., those of Horowitz et al. \cite{horo01a, 
horo01b,horo06}, study the possible nontrivial effect of 
introducing a second alternate position of sticking 
(see Fig \ref{horoBD},\ref{horoBD1},\ref{horoBD2}). Each incoming particle 
either deposits on top of a selected site as in RD, or, sticks to 
a higher location of a taller nearest neighbour. 
The probability assigned is ad hoc and is not suggested by any 
underlying mechanism or dynamics. 
The GBD, on the other hand, represents a true stochastic 
process. The incoming particle, at each contact with the surface 
has a chance to stick (with probability $p$) or to slide 
down (with probability $q=1-p$). This process continues for successive 
points of contact until the descending particle reaches the 
bottom of the column. The case of particle descending between two 
nearest neighbor adjacent columns is also considered, where the probabilities of
sticking to two adjacent surfaces of contact is different than that for a single
nearest neighbor column (see Fig. \ref{gRDBD}).

In some other competitive growth models, different species of
particles are considered, some deposit as in RD, others as in BD.
These models use mixture of particles with different pre-assigned 
probability of sticking. Thus, they represent deposition of mixed 
species of particles.
However, no individual particle has the possibility of sticking
to successive contact locations guided by a relevant stochastic process.
Thus, in these other varaints of BD and competitive growth models, 
the possible sticking positions of the new particle are 
far fewer than those in the GBD model proposed in this work.
The configurations possible in the 
competitive growth models (Fig \ref{horoBD},\ref{horoBD1},\ref{horoBD2})
form a smaller subset of the large 
ensemble of configurations allowed in the present
model(Fig \ref{gRDBD}).

The present GBD model involves stochasticity at two stages, first in the
random selection of an active site and second, in
the assignment of a sticking probability of the incoming particle .
The structure of the deposit in the GBD model, depends on 
the successive stages of evolution. Hence, the relevant evolution
equation is expected to be a stochastic integro-differential equation
(SIDE) \cite{katzav2003, kas14}. This is in contrast to Edwards-Wilkinson 
(EW) or Kardar-Parisi-Zhang (KPZ) equations \cite{ew82,kpz86} which 
are stochastic differential equation (SDEs).


\section{Generalized Ballistic Deposition Model}\label{sec2}
The generalized ballistic deposition model presented in this work 
attempts to represent the deposition of realistic particles with 
varying degrees of stickiness.
A parameter $p$, whose values vary between 0 and 1, is introduced
that represents the probability of an incoming particle sticking
to a point of contact on the growing surface.
A particle is allowed to descend vertically towards a randomly 
chosen site on a one dimensional substrate. If the selected site is 
higher than its nearest neighbors,
the particle simply deposits on top of the column at that site.
However, if the 
chosen site has a taller column of particles as its nearest neighbor, then 
the new particle sticks to the first occupied site it encounters if
the value of $p$ is larger than a random number generated 
from a uniform distribution between 0 and 1. Otherwise, it slides
down vertically to the next occupied site with probability $(1-p)$.
At this site the particle may stick with probability $p(1-p)$
or continue its further descent with probability $(1-p)^2$, and 
so on, till it reaches the bottom.
Thus if the chosen site has a nearest neighbor with column height taller
by $n$ layers relative to it, the probabilities of the arriving 
particle sticking to the successive particles of the nearest neighbor 
column from top are given by,
\beq
P(1) = p, \; P(2) = p(1-p), \dots \; P(k) = p(1-p)^{(k-1)}.
\eeq
The probability that the particle slides past the preceding $(n-1)$
occupied neighbors, and lands at the lowest possible position is given by,
\bd
P(n) = 1 - \sum_{k=1}^{n-1} P(k) = (1 - p)^{(n-1)}.
\ed
This describes a proper stochastic process. The total probability of a 
descending particle sticking to one of the allowed position is 
$\sum_{k = 1}^{n} P(k) = 1$.

In simple deposition models, the surface width follows a dynamic scaling law
\cite{fam86},
\beq
W(L,t) \sim  L^{\alpha}f(t/L^z),
\eeq
where $f$ is a scaling function satisfying
$f(\infty) \sim \;constant$ and $f(x) \sim x^{\beta}$  for small $x$.
The exponents $\alpha, \beta$ and $z$ are related by $z = \alpha/\beta$.

For the GBD studied in this work,
the introduction of a sticking probability $p$, brings in 
another parameter in the problem. GBD interpolates between
RD ($p=0$) and BD ($p=1$) systems. 
Physically relevant quantities, e.g., surface width $W(L, p, t)$
and porosity $\rho (L, p, t)$ thus depend on the sticking probability $p$
in addition to $L$ and $t$.
From the results of our simulation we obtain dynamic scaling relations,
\beq\label{wscale}
W(L,p,t) \sim L^{\alpha}p^{-\alpha'}F\left(\frac{t \, p^{z'}}{L^z}\right) 
\eeq
\beq\label{poroscale}
\rho(L,p,t) \sim L^a p^b G\left(\frac{t\,p^{d}}{L^c}\right) .
\eeq
where $F(x)$ and $G(y)$ are scaling functions described above.

\section{Results and discussion}
\begin{figure}[!ht]
  {\includegraphics[width=0.95\linewidth]{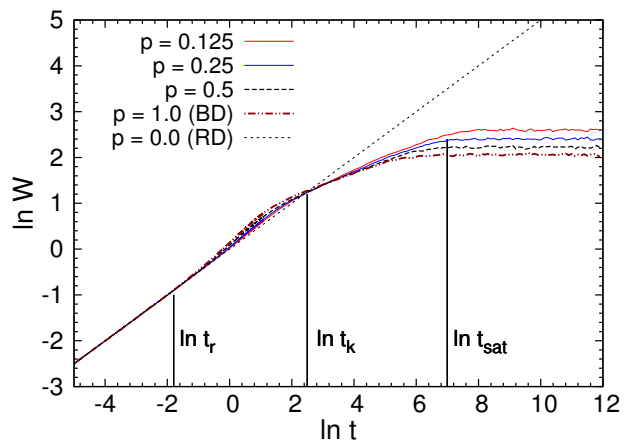}}
  \caption{(Color Online) Logarithmic plot of surface
width with time for different $p$ and system size $L=256$, showing four distinct regions and 
crossover times.\label{lnwvslnt_L}}
\end{figure}

For different values of sticking probabilities between $p=0$ and $p=1$,
simulations were performed in $(1+1)$ dimension for system sizes 
$L=16, 32, 64, 128, 256, 512$ and $1024$. The graphs are suitably drawn to 
present the data, observations and results in a succinct, yet clear 
and uncluttered manner.
The average number of layers deposited, is used as a measure of time $t$.
Depending on the values of $p$ and $L$, the simulation results were averaged
over 1000 to 5000 ensembles.
The logarithmic plot of surface width with time has  
four distinct regions, for any non-zero probability ($p>0$). This
is shown in Fig \ref{lnwvslnt_L}.

The dependence of surface width $W$ on $t$ in log-log scale,
in the early submonolayer region ($t \ll 1$) is linear 
with slope $1/2$ as in random deposition (growth region 1, GR-1). 
At later stages of submonolayer growth (growth region 2, GR-2), $t \simeq 1^-$, the
surface width shows a steep increase which continues for the 
first few layers ($1-\epsilon \le t\le 3$, $ 1 \gg \epsilon > 0$). 
With deposition of further layers, 
the rate of increase in width slows down (growth region 3, GR-3). 
After deposition of a large number of layers, the ensemble average of the surface width 
saturates. Three different crossover times are of 
relevance. The first crossover time $t_r$ corresponds to the change 
from random growth to region with slope greater than $1/2$. 
The second crossover time $t_k$ corresponds to time beyond few layers 
where the slope decreases and changes from GR-2 to GR-3. 
The third crossover time $t_{sat}$ corresponds to beginning 
of saturation region.

\begin{figure}[!ht]
  {\includegraphics[width=0.95\linewidth]{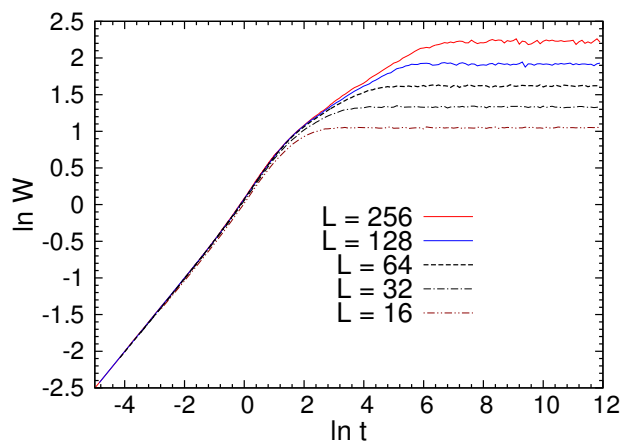}}
  \caption{(Color Online) Variation of $\ln W$ with  $\ln t$ for $p=0.5$
for different system sizes.  \label{lnwvslnt_p}}
\end{figure}

The appearance of different growth regions in the present
model may be understood as follows.
Initially, when the deposition starts from a flat substrate almost no two adjacent sites are occupied.
Hence there is no correlation among neighboring columns and 
the growth is random like, irrespective of whether the model allows for sticking or not.
This feature is observed in all systems with
different $L$ and $p$ values as shown in Fig \ref{lnwvslnt_L} and 
Fig \ref{lnwvslnt_p}. 

\begin{figure}[!htb]
  {\includegraphics[width=0.95\linewidth]{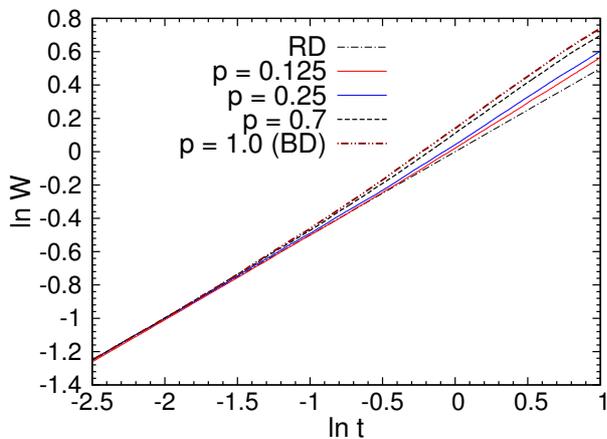}}
  \caption{(Color Online) Deviation from random deposition behavior at later stages of submonolayer growth for $L=256$. \label{devrandom}}
\end{figure}

\begin{figure}[!ht]
  {\includegraphics[width=0.95\linewidth]{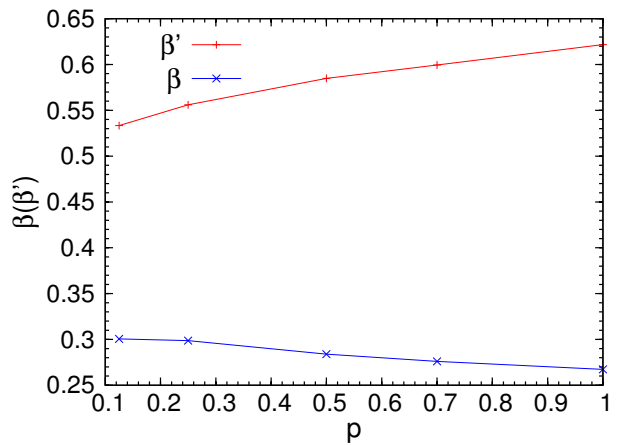}}
  \caption{(Color Online) Dependence of $\beta^{\prime}$ and $\beta$ on $p$ shown for system size $L=512$. \label{bbprime}}
\end{figure}

The deviation from random like behavior in the
few layer deposition region is shown in Fig \ref{devrandom}. 
The rate of growth of surface width 
in this region is higher than that in the case of random deposition.
As the number of particles deposited is nearly $L$, due to fluctuation, 
some short multi-layer columns may start forming.
Hence the descending particles encounter occupied neighbors,
and allowing sticking in the model, brings in non-trivial correlations in the system.
However, since very few layers are deposited at this stage, even one
particle sticking to a higher location or descending to the bottom of a
column, makes a significant relative change in width.
Thus the rate of growth of surface width in this region,
when only a few layers have formed, is higher than that for RD.
Our study shows, that the growth exponent in this region, denoted by $\beta^{\prime}$,
increases with sticking probability $p$, as is shown in Fig \ref{bbprime},
reaching a maximum when $p=1.0$, corresponding to the standard BD,
and is almost independent of system size $L$, as is shown in 
Fig \ref{lnwvslnt_p}.

The rate of increase in surface width slows down as more and more particles are
deposited in GR-3. The average height of the interface and its width are larger. 
The descending particle need not proceed to the bottom, and can get deposited
at a higher location by sticking. Thus the correlation has a smoothening effect,
as it fills up deep crevices efficiently.
The growth exponent $\beta$ in this region decreases with 
$p$, unlike the exponent in GR-2.
With further deposition of particles the surface width finally saturates.
The saturated width depends both on the system size $L$ and sticking probability $p$,
as shown in Fig \ref{lnwvslnt_L} and Fig \ref{lnwvslnt_p}.

For a given value of $p$, the surface width at saturation $W_{sat}$, and the time
at which the saturation is reached $t_{sat}$,
increase with system size $L$.
For a given system size $L$, $W_{sat}$ and $t_{sat}$ decrease with increase in 
probability of sticking $p$. This decrease is more pronounced for lower values of $p$.

\begin{figure}[!ht]
  {\includegraphics[width=0.95\linewidth]{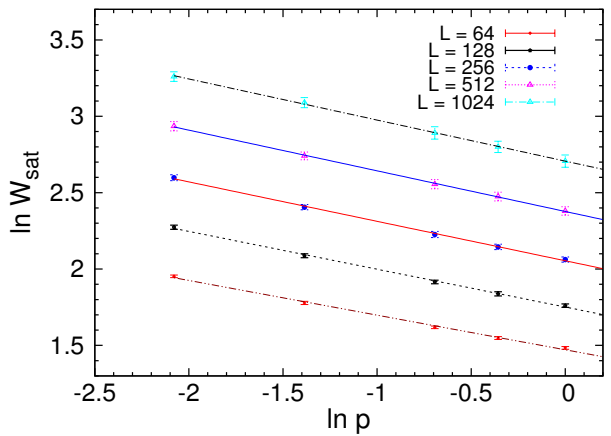}}
  \caption{(Color Online) Variation of $\ln W_{sat}$ with
$\ln p$ for different system sizes $L$.
\label{lnwslnL}}
\end{figure}

The dependence of $W_{sat}$ on $p$, for a given system size, 
is of the form  $W_{sat}(L,p) \simeq f(L) \cdot p^{- \alpha'}$, where
the exponent $\alpha'$ is independent of $L$ (Fig \ref{lnwslnL}).
An increase in sticking probability $p$ corresponds to 
a stronger correlation in the deposition process. The surface width
saturates at lower values of saturation width $W_{sat}$, at corresponding
earlier times, i.e., smaller $t_{sat}$.

\begin{figure}[!ht]
  {\includegraphics[width=0.95\linewidth]{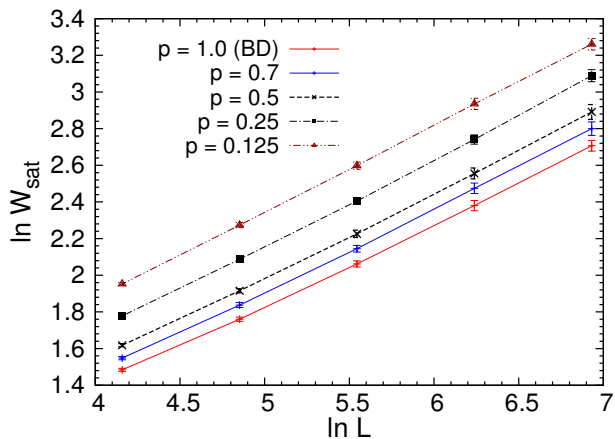}}
  \caption{ (Color Online) Variation of the saturated surface width $\ln W_{sat}$ with 
$\ln L$ for different $p$ values. \label{lnwslnp}}
\end{figure}

\begin{figure}[!ht]
  {\includegraphics[width=0.95\linewidth]{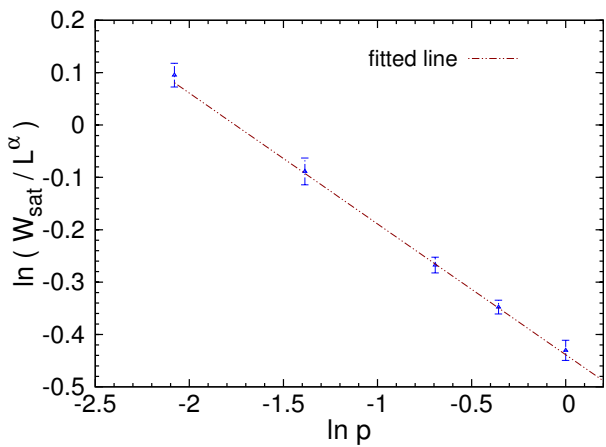}}
  \caption{ (Color Online) Scaled saturated surface width $\ln (W_{sat}/L^{\alpha})$ versus  $\ln p$. 
  The slope is $- 0.250$.
  \label{wsatscale}}
\end{figure}

For a given probability of sticking $p$, $\ln W_{sat}$ increases
linearly with $\ln L$. The dependence is found to be of the form
$W_{sat}(L,p) \sim L^{\alpha}$ (Fig \ref{lnwslnp}). 
From results of extensive simulations, graphically presented in the adjacent figures, Fig \ref{lnwslnL}, Fig \ref{lnwslnp},
and Fig \ref{wsatscale},
we find the exponents $\alpha = 0.452$ and $\alpha'= 0.250$,
in the region of saturated surface width.

\begin{figure}[!ht]
  {\includegraphics[width=0.95\linewidth]{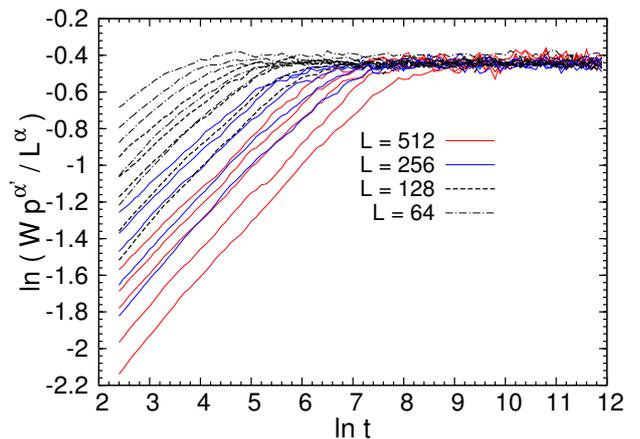}}
  \caption{(Color Online) Variation of $\ln (W \, p^{\alpha'} / L^{\alpha})$ with $\ln t$ showing 
dependence on system size $L$ and sticking probability $p$. 
Within each set of $L$, the different $p$ lines appear in order of 
decreasing $p$ values as one moves from bottom to top of the plot.\label{yfullscale}}
\end{figure}

\begin{figure}[!ht]
  {\includegraphics[width=0.95\linewidth]{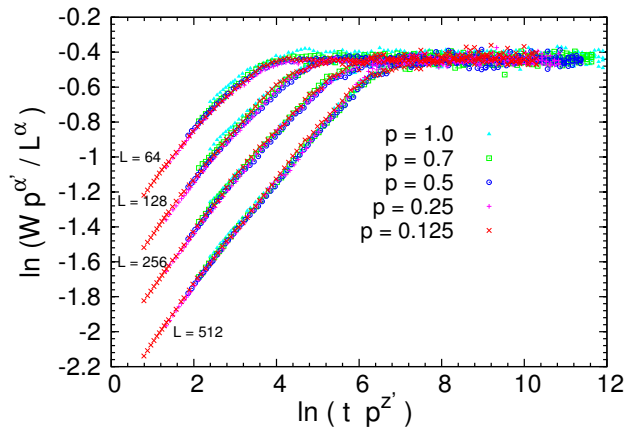}}
  \caption{ (Color Online) $\ln (W\,p^{\alpha'}/L^{\alpha})$ versus $\ln (t\,p^{z'})$ for different sticking probability $p$. \label{Wfulltp}}
\end{figure}

\begin{figure}[ht!] 
  {\includegraphics[width=0.95\linewidth]{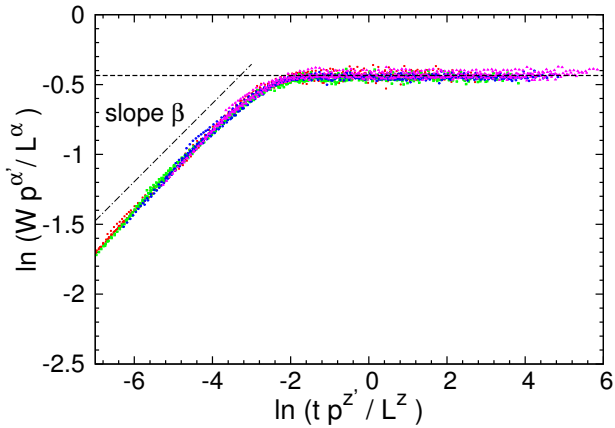}}
  \caption{(Color Online) $\ln (W\,p^{\alpha'}/L^{\alpha})$ versus $\ln (t\,p^{z'}/L^{z})$ 
for $L = 512,256,128,64$ and $p = 1.0,0.7,0.5,0.25,0.125$. \label{fscale}}
\end{figure}

The surface width $W$ in the third growth region, GR-3, i.e., between $t_k$ and
$t_{sat}$ depends on both system size $L$ and sticking 
probability $p$ as shown in Fig \ref{yfullscale}.
The scaled width $( W p^{\alpha'}/L^{\alpha} )$ in GR-3 
shows that it is larger for larger system sizes and lower probability ($p=0.125$). 
For a given system size $L$, $t_{sat}$ and hence 
the growth region, decrease with increase in probability $p$.
The scaling of $t$ with respect to $p$ is obtained from 
Fig \ref{yfullscale}, with exponent $z^{\prime} = 0.7722$.
Log-log plot of rescaled variables
$( W p^{\alpha'}/L^{\alpha} )$ versus $(t\, p^{z'}/L^z)$
shows an excellent collapse of data in the growth region GR-3 and saturation region as
shown in Fig \ref{Wfulltp} and Fig \ref{fscale}.

Analysis of our simulation data produces numerical estimates for the scaling exponents
in Eq.\ref{wscale}. The exponents are,
 $\alpha = 0.452 \pm 0.016$, $\alpha^{\prime} = 0.250 \pm 0.016$, $z = 1.45 \pm 0.22$ 
and $z^{\prime} = 0.77 \pm 0.04$.
The closest representation of these values in terms of rational fractions are 
$\alpha = 1/2$, $\alpha^{\prime} = 1/4$, $z = 3/2$ and $z^{\prime} = 3/4$.
From the scaled $W$ versus scaled $t$ graph in Fig.\ref{Wfulltp} and Fig.\ref{fscale},
we obtain $\beta \simeq .312$ as the slope in the growth region GR-3. 
In Eq. \ref{wscale},
for small values of $(t p^{z'}/L^z)$,
we can approximate $F(x) \simeq x^{\beta}$, and hence
\beq
W \sim L^{\alpha} p^{-\alpha'} \, \left(\frac{t p^{z'}}{L^z}\right)^{\beta},
\eeq
giving $\beta$ as the exponent of $t$ in this regime.
The exponents obtained above, satisfy the relations,
\beq
\beta = \frac{\alpha}{z}=\frac{\alpha'}{z'}.
\eeq
The growth exponent $\beta$ is approximately the rational fraction $1/3$.

The porosity $\rho$ can be defined as the fraction of unoccupied sites 
within a few layers $N$ just below the active interface, where no further
deposition can take place. 
It is seen that the porosity
is quite independent when $N$ is varied from $16$ to $L$ for a given system size $L$.
The porosity is found to initially increase with
time, signifying growth and then saturates to a value $\rho_{sat}$.
The onset of saturation for porosity occurs earlier than 
the onset of
saturation of surface width. 

\begin{figure}[ht!] 
  {\includegraphics[width=0.95\linewidth]{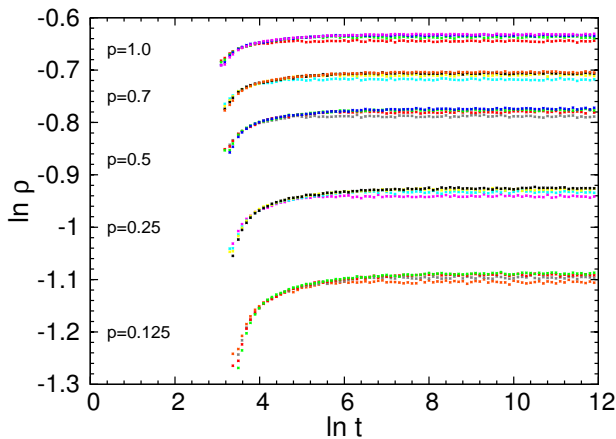}
  \caption{ (Color Online)  $\ln \rho$ versus $\ln t$
    for $L = 512,256,128,64$ and $p = 1.0,0.7,0.5,0.25,0.125$.\label{punscale}}}
\end{figure}

\begin{figure}[ht!] 
  {\includegraphics[width=0.95\linewidth]{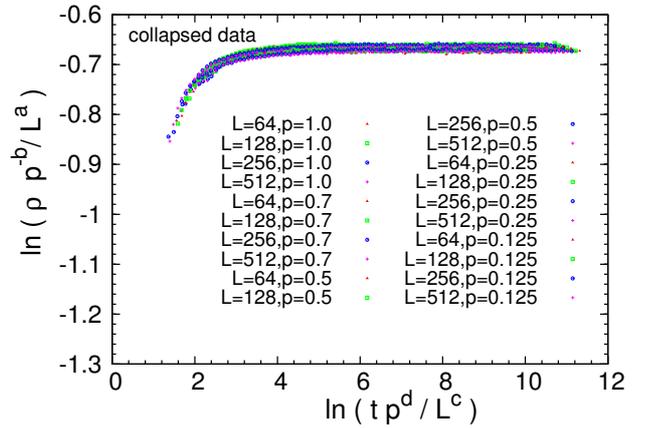}
  \caption{ (Color Online)  Scaled data shown as $\ln (\rho \, p^{-b}/L^a)$ versus $\ln (t\,p^d/L^c)$
    for $L = 512,256,128,64$ and $p = 1.0,0.7,0.5,0.25,0.125$.\label{pscale}}}
\end{figure}

The porosity  $\rho$ is found to depend on the sticking probability $p$ and shows a
weak dependence on the system size $L$.
$\rho_{sat}$ is higher for higher $L$ at any given $p$.
However, in the early growth region, $\rho(L,p,t)$ starts at a lower
value for higher $L$.
The porosity in both the growth and saturation regions increases 
with increase in sticking probability $p$; fully ballistic deposition system
being most porous.
Onset of saturation in porosity is earlier for higher sticking probability.
A scaling behavior for porosity is obtained in the form of Eq.\ref{poroscale},
with the exponents $a = 0.00690 \pm 0.00070, b = 0.2204 \pm 0.0051, c = 0.134 \pm 0.035$ and
$d = 0.59 \pm 0.16$.
The dependence of porosity on time $t$, sticking probability $p$ and system size $L$
is shown in Fig \ref{punscale} and as scaled data in
Fig \ref{pscale}.

It may be noted that, in our simulations of the present 
generalized ballistic deposition model,
two independent random number generators were
used, one for selecting a site on the growing surface and,
another to determine whether a particle will stick
at a particular location for a chosen value of the sticking probability.
These two random number generators are completely independent 
and uncorrelated to each other.

\section{Conclusion}
We have studied the deposition of physically realistic particles 
with variable 
stickiness by proposing a generalized ballistic model of deposition.

As mentioned in section \ref{sec2}, the GBD model involves 
stochasticity in two stages:(i) in random selection of an active site and 
(ii) in assigning a sticking probability that affects 
the extent to which the height at the active site is altered for
a given configuration of particles deposited on neighboring sites.
The random selection of an active site appears
as the random noise term in the corresponding differential 
equation, such as in the KPZ equation, while the detailed 
mechanism of the change in height $h(x, t)$, determines
the different spatial derivatives i.e., slope, curvature etc.,
appearing in the differential equations.

In the present model, the sticking of an incoming particle 
at a site is possible only when it has slipped past the 
surfaces encountered earlier in its path.
Thus, the possibility of sticking at
the present position, depends on the probabilities of all
such earlier events. The relevant equation and the scaling
behavior is expected to depend on this sticking probability.
This is corroborated by our observations.

A larger value of sticking probability implies stronger
correlation among neighboring columns, hence, the surface width,
at a fixed value of $L$, is expected to saturate 
to lower values of $W_{sat}$ at earlier (smaller $t_{sat}$) times.
In the growth region, however, correlations have a 
smoothening effect, as deep creivces are filled up more efficiently.
Thus, in the growth region, surface width, at a fixed $L$,
should decrease with increase in sticking probability.
Porosity, on the other hand, should increase
with the sticking probability, in both growth and 
saturation regions for a fixed system size, as a 
larger sticking probability implies a more porous structure.
Our observations are in agreement with the above arguments.

We find excellent collapse of data for
the scaling of the surface width and porosity, in terms of system size
$L$ and sticking probability $p$ in the KPZ-like growth and
saturation regions. 

A simplying assumption in this work is that the sticking 
probability at successive encounters is assumed to be
a constant. However, each encounter may cause some 
physical change(s) to the incoming particle. Further 
investigation of the above may be of interest and progress
of study along the above lines will be reported elsewhere.
It may be noted however, that a most general treatment
will involve all the complexities associated with 
non-Markovian processes.

\section{Acknowledgement}
Authors gratefully acknowledge the use of computing facility at University of
Calcutta, under the DRS program sponsored by the University Grants Commission, India.
B. M. also wishes to acknowledge the financial assistance provided by the West Bengal State Departmental Fellowship, India.

\bibliographystyle{apsrev4-1}
\bibliography{myref}
 
\end{document}